\documentclass[aps,prc,preprint,superscriptaddress,showpacs,floatfix]{revtex4}

\usepackage{mathrsfs}
\usepackage{latexsym}
\usepackage{amsmath}
\usepackage{amssymb}
\usepackage{graphicx}
\usepackage{longtable}
\usepackage{multirow}
\usepackage{bbm}
\usepackage{epsfig}
\usepackage[usenames]{color}
\usepackage[colorlinks=true,citecolor=blue,linkcolor=blue]{hyperref}
\allowdisplaybreaks

\begin{document}



\title{Baryon number fluctuations induced by hadronic interactions at low temperature and large chemical potential}

\author{Guo-yun Shao}
\email[Corresponding author: ]{gyshao@mail.xjtu.edu.cn}
\affiliation{School of Science, Xi'an Jiaotong University, Xi'an, 710049, China}
\affiliation{ MOE Key Laboratory for Nonequilibrium Synthesis and Modulation of Condensed Matter, Xi'an Jiaotong University, Xi'an, 710049, China}

\author{Xue-yan Gao}
\affiliation{School of Science, Xi'an Jiaotong University, Xi'an, 710049, China}

\author{Wei-bo He} 
\affiliation{School of Science, Xi'an Jiaotong University, Xi'an, 710049, China}

\begin{abstract}
 
 We investigate the density fluctuations induced by  hadronic interactions at low temperatures and large chemical potentials after the hadronization of quark-gluon plasma~(QGP). We analyze the structures of net-baryon number kurtosis and skewness and elaborate on their relations with the nuclear liquid-gas~(LG) phase transition and  hadronic interactions above the critical temperature. Combining with the relevant experimental projects at RHIC/NICA/FAIR/J-PARC/HIAF with collision energies from $200\,$GeV  to $1.8\,$GeV, we propose a double-peak structure of kurtosis  (skewness) as a function of collision energy, which can be taken to identify the chiral phase transition  and the nuclear liquid-gas phase transition. In particular, the fluctuation distributions at low temperatures provide a new method to explore  hadronic interactions and nuclear liquid-gas transition. 
\end{abstract}


\maketitle

\section{introduction}
The exploration of QCD phase structure and the search for phase transition signals  are hot topics in both theory and heavy-ion collision experiments. 
The smooth crossover essence from 
quark-gluon plasma~(QGP) to hadrons at small chemical potential and high temperature has been indicated in lattice QCD~(LQCD) simulations~\cite{Aoki06, Bazavov12, Borsanyi13,Bazavov14, Bazavov17,Borsanyi14}. However, the calculations based on effective quark models~(e.g., \cite{Fukushima04,Ratti06,Costa10,Fu08, Sasaki12, Ferreira14, Schaefer10, Skokov11}), Dyson-Schwinger equation approach~\cite{Qin11,Gao16, Fischer14,Shi14} and the functional renormalization group theory~\cite{Fu19} 
show that the chiral condensate undergoes a first-order  transition at  large chemical potential.  At the end of the first-order transition line there exists a critical end point~(CEP) connecting with the chiral crossover separation line.

To extract phase transition signals in experiments, the high-order cumulants of net proton~(proxy for baryon) have been measured in the first phase of Beam Energy Scan (BES-I) at RHIC STAR~\cite{Aggarwal10, Adamczyk14}, and a non-monotonic energy dependence of  the net-proton number kurtosis  $\kappa \sigma^2$ was discovered~\cite{Luo2014, Luo2017}.  The $\kappa \sigma^2$ shows a large deviation from the Poisson baseline and the prediction of the hadron resonance gas~(HRG) model. These behaviors of net-proton number kurtosis  possibly hint that the STAR experiments in Au+Au collisions with $\sqrt {s_{NN}}=7.7\sim27\,$GeV pass through the QCD critical region. 
To improve the measurement accuracy and confirm the non-monotonic energy dependence of net-proton number kurtosis, RHIC has already started BES-II program.  The fixed-target mode experiment with lower collision energies is also in plan~\cite{Luo2017}. There are also relevant experimental projects at NICA/FAIR/J-PARC/HIAF  to search for the location of  CEP and QCD phase boundary. 

The chemical freeze-out conditions can be extracted from the experimental data in the thermal statistical model~\cite{Andronic18}.  
The results show that with the decrease of colliding energy to several GeV the generated strongly interacting matter has large chemical potentials and low temperatures after hadronization. 
It means that high-density nuclear matter possibly forms at low temperatures with the decrease of collision energy. Therefore, in the low-temperature region the interactions of nuclear matter play an important role on the density fluctuations. 
It is also known that the nuclear liquid-gas transition exists at low temperatures~\cite{Chomaz04,Pochodzalla95,Borderie01,Botvina95,Agostino99,Srivastava02,Elliott02}. Similar to the chiral transition in the critical region, the fluctuations at low temperatures near the LG phase transition is also possibly strong.  Even in the region not very close to the nuclear LG transition, the interaction between hadrons will affect the density fluctuation distributions.

Considering the future HIC experiments  with the decreased collision energies, in this study we will investigate how hadronic interactions affect the density fluctuations at low temperatures and large chemical potentials, and compare the phase transition of hadronic matter with the quark chiral transition. We will also discuss possible signals in experiments with 
different collision energies to explore the phase structure of strongly interacting matter.

The paper is organized as follows. In Sec.~II, we briefly introduce the quantum hadron dynamics~(QHD) model and the formulas to describe fluctuations of conserved charges. In Sec.~III, we illustrate the numerical results of net-baryon number fluctuations, and discuss the relations with hadronic interactions. We also compare the results with the fluctuations of quark chiral transition, and analyze the possible signals in experiments. A summary is finally given in Sec. IV.

\section{the nonlinear Walecka model and fluctuations of conserved charges}

It is still lack of effective methods to describe the properties of hadronic matter based on QCD theory, therefore, we take the quantum hadron dynamics  model  to describe the hadronic matter after the hadronization of QGP at low temperatures.
This model describes well the properties of finite nuclei and nuclear matter.  The study in \cite{Fukushima15} also indicates its approximate equivalence to the hadron resonance gas model at low temperature and small density.

For nucleons-meson system, the Lagrangian density in the nonlinear Walecka model \cite{Glendenning97} is 
\begin{eqnarray}\label{lagrangian}
\cal{L}\!&\!=\!&\sum_N\bar{\psi}_N\!\big[i\gamma_{\mu}\partial^{\mu}\!-\!(\!m_N
          \!-\! g_{\sigma }\sigma)\!
                 \! -\!g_{\omega }\gamma_{\mu}\omega^{\mu} \!-\! g_{\rho }\!\gamma_{\mu}\boldsymbol\tau_{}\!\cdot\!\boldsymbol
\rho^{\mu} \big]\!\psi_N       \nonumber\\
         & &    +\frac{1}{2}\left(\partial_{\mu}\sigma\partial^
{\mu}\sigma-m_{\sigma}^{2}\sigma^{2}\right)\!-\! \frac{1}{3} bm_N\,(g_{\sigma} \sigma)^3-\frac{1}{4} c\,
(g_{\sigma} \sigma)^4
                    \nonumber\\
       & &+\frac{1}{2}m^{2}_{\omega} \omega_{\mu}\omega^{\mu}
          -\frac{1}{4}\omega_{\mu\nu}\omega^{\mu\nu}  \nonumber \\
& &    +\frac{1}{2}m^{2}_{\rho}\boldsymbol\rho_{\mu} \! \cdot \! \boldsymbol
\rho^{\mu}   \!- \! \frac{1}{4}\boldsymbol\rho_{\mu\nu} \! \cdot \! \boldsymbol\rho^{\mu\nu} ,
 \end{eqnarray}
 where $
\omega_{\mu\nu}= \partial_\mu \omega_\nu - \partial_\nu
\omega_\mu$, $ \rho_{\mu\nu} =\partial_\mu
\boldsymbol\rho_\nu -\partial_\nu \boldsymbol\rho_\mu$. The interactions between
baryons are mediated by $\sigma,\,\omega,\,\rho$ mesons. 
The model parameters, $g_\sigma, g_\omega, g_\rho, b$ and $ c$, are fitted with the compression modulus $K=240\,$MeV, the symmetric energy  $a_{sym}=31.3\,$MeV, 
the effective nucleon mass $m^*_N=m_N- g_\sigma \sigma=0.75m_N$~($m_N$ is the nucleon mass in vacuum) and the
 binding energy $B/A=-16.0\,$MeV at nuclear saturation density with $\rho_0=0.16\, fm^{-3}$.

 The thermodynamical  potential of the nucleons-meson system can be derived in the mean-field approximation as
\begin{eqnarray}
 \Omega\!&\!=&\!-\!\beta^{-1} \sum_{N} 2\! \int \frac{d^{3} \boldsymbol{k}}{(2 \pi)^{3}}\!\bigg[\ln \!\left(1+e^{-\beta\left(E_{N}^{*}(k)-\mu_{N}^{*}\!\right)}\right)\! \nonumber \\ 
 &&+\!\ln\! \left(1\!+\!e^{-\beta\left(E_{N}^{*}(k)+\mu_{N}^{*}\right)}\right)\!\bigg]\! +\!\frac{1}{2} m_{\sigma}^{2} \sigma^{2}\!+\!\frac{1}{3} b m_{N}\left(g_{\sigma} \sigma\right)^{3} \nonumber \\
 &&+\frac{1}{4} c\left(g_{\sigma} \sigma\right)^{4}-\frac{1}{2} m_{\omega}^{2} \omega_{}^{2}-\frac{1}{2} m_{\rho}^{2} \rho_{3}^{2}  ,
  \end{eqnarray}
where $\beta=1/T$, $E_{N}^{*}=\sqrt{k^{2}+m_{N}^{*2}}$,  and  $\rho_{3}$ is the third component of $\rho$ meson field.  The effective chemical potential $\mu_{N}^{*}$ is defined as  $\mu_{N}^{*}=\mu_{N}-g_{\omega} \omega_{}-\tau_{3 N} g_{\rho} \rho_{3}$ ($\tau_{3 N}=1/2$ for proton, $-1/2$ for neutron).

By minimizing the thermodynamical potential
\begin{equation}
\frac{\partial \Omega}{\partial \sigma}=\frac{\partial \Omega}{\partial \omega_{}}=\frac{\partial \Omega}{\partial \rho_{3}}=0,
\end{equation}
the meson field equations can be derived as
\begin{equation}\label{sigma}
g_{\sigma} \sigma\!=\!\left(\frac{g_{\sigma}}{m_{\sigma}}\right)^{2}\!\left[\rho_{p}^{s}+\rho_{n}^{s}-b m_{N}\left(g_{\sigma} \sigma\right)^{2}-c\left(g_{\sigma} \sigma\right)^{3}\right],
\end{equation}
\begin{equation}
g_{\omega} \omega=\left(\frac{g_{\omega}}{m_{\omega}}\right)^{2}\left(\rho_{p}+\rho_{n}\right),
\end{equation}
\begin{equation}\label{rho}
g_{\rho} \rho_{3}=\frac{1}{2}\left(\frac{g_{\rho}}{m_{\rho}}\right)^{2} \left(\rho_{p}-\rho_{n}\right).
\end{equation}

In the eqs.(\ref{sigma})-(\ref{rho}), the nucleon number density
\begin{equation}
\rho_N\!=\!2\! \int\! \frac{d^{3} \boldsymbol{k}}{(2 \pi)^{3}} [f\left(E_{N}^{*}(k)-\mu_{N}^{*}\right)\!-\!\bar{f}\left(E_{N}^{*}(k)+\mu_{N}^{*}\right) ],
\end{equation}
and the scalar density
\begin{eqnarray}
\rho_{N}^{s}&=&2 \int \frac{d^{3} \boldsymbol{k}}{(2 \pi)^{3}} \frac{m_{N}^{*}}{E_{N}^{*}(k)}[f\left(E_{N}^{*}(k)-\mu_{N}^{*}\right)  \nonumber \\
& & \hspace{80pt}   +\bar{f}\left(E_{N}^{*}(k)+\mu_{N}^{*}\right)],
\end{eqnarray}
where $f(E_{N}^{*}(k)-\mu_{N}^{*})$ and $\bar{ f} (E_{N}^{*}(k)+\mu_{N}^{*})$ are the  fermion and antifermion distribution functions.

For a given temperature and chemical potential (or baryon number density), the meson field equations can be solved.  In the present study,   
symmetric nuclear matter is considered to give an essential description of the density fluctuations.

For a thermal medium, the fluctuations of conserved charges are most sensitive observables to study the phase transition, in particular, the critical phenomenon.  
The pressure of the system for a grand-canonical ensemble is related to the logarithm of the partition function \cite{Karsch15}:
\begin{equation}\label{}
\frac{P}{T^4}=\frac{1}{VT^3}\ln[Z(V,T,\mu_B, \mu_Q, \mu_S)],
\end{equation}  
where $\mu_B, \mu_Q, \mu_S$ are the chemical potentials of conserved charges, the baryon number, electric charge and strangeness, respectively. The generalized susceptibilities can be derived by taking the partial derivatives of the pressure with respect to the corresponding chemical potentials \cite{Luo2017}
\begin{equation}\label{}
\chi^{BQS}_{ijk}=\frac{\partial^{i+j+k}[P/T^4]}{\partial(\mu_B/T)^i \partial(\mu_Q/T)^j \partial(\mu_S/T)^k},
\end{equation}  
The cumulants of multiplicity distributions of the conserved charges are connected with the generalized susceptibilities by
 \begin{equation}\label{}
\!C^{BQS}_{ijk}\!=\!\frac{\partial^{i+j+k}\ln[Z(V,T,\mu_B, \mu_Q, \mu_S)]}{\partial(\mu_B/T)^i \partial(\mu_Q/T)^j \partial(\mu_S/T)^k}\!=\!VT^3\chi^{BQS}_{ijk}\, .
\end{equation} 
In experiments, observables are constructed by the ratio of cumulants, which cancel the volume dependence and then can be compared with theoretical calculations of the generalized susceptibilities.

To compare with the observables in experiments, we focus on two important statistic distributions, the skewness ($S\sigma$) and kurtosis ($\kappa \sigma^2$) of  net-baryon number fluctuations. The  $S\sigma$ and  $\kappa \sigma^2$ are related to the high order cumulants and susceptibilities as
\begin{equation}\label{comulants}
S\sigma=\frac{C^B_3}{C^B_2}=\frac{\chi^B_3}{\chi^B_2} \,\,\,\,\,\texttt{and}\,\,\,\,  \kappa \sigma^2=\frac{C^B_4}{C^B_2}=\frac{\chi^B_4}{\chi^B_2}.
\end{equation}

\section{numerical results}

\begin{figure}[htbp]
\begin{center}
\includegraphics[scale=0.3]{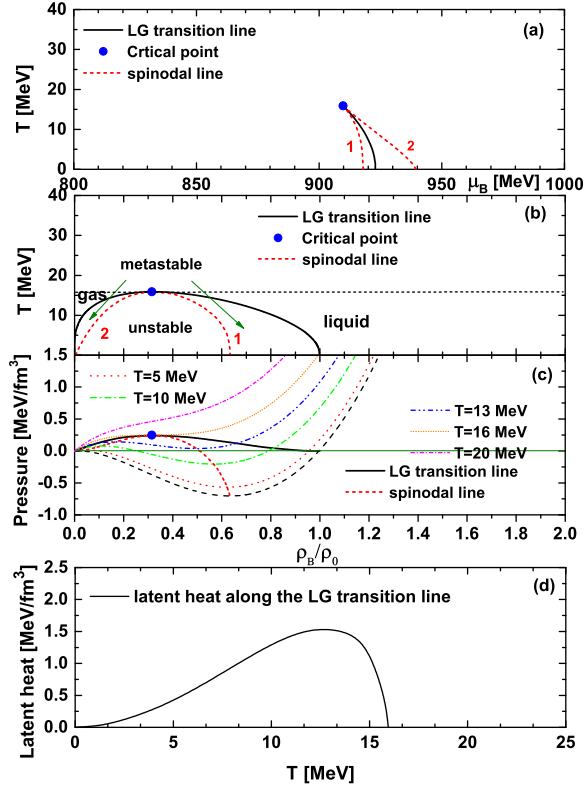}
\caption{\label{fig:1}Panel (a): nuclear LG phase transition line and the boundaries of  the spinodal region in the $T-\mu_B$ plane. Panel (b):~nuclear LG transition and the spinodal instability region in $T-\rho_B$. Panel (c): The equation of state of hadronic matter at different temperatures.  Panel (d): The latent heat along the liquid-gas transition line.
}
\end{center}
\end{figure}

To better understand the relation between net-baryon number fluctuations and the phase transition of hadronic matter, we first plot in Fig.~\ref{fig:1} the phase diagram of the nuclear LG transition at low temperature. 

 In the panel (a), the solid line is the first-order transition line, and the blue solid dot is the critical end point.  The red dashed lines are the boundaries of the spinodal region. A similar first-order transition is derived in modified HRG model with the inclusion of van der Waals Interactions~\cite{Vovchenko17}. In the panel (b), we plot the the liquid-gas transition in the $T-\rho_B$ plane. The black solid lines on the two sides of the critical point correspond to the first-order transition. The region between the red dashed lines presents the spinodal instability region. In the panel (c), we plot the pressure of hadronic matter as a function of density for different temperatures.  For a given temperature, the spinodal region~(between the red dashed lines) can be derived with the mechanical instability condition $\partial P/ \partial \rho_B<0$, and the first-order transition~(black solid lines) can be derived in the stable region with the Gibbs conditions. In the panel (d), we show the latent heat along the nuclear liquid-gas transition line. We note that the latent heat becomes zero at the critical point.

As shown in Panel (a) of Fig.~\ref{fig:1}, the first-order nuclear phase transition is derived according to the conditions of two-phase equilibrium.
In the following, we will supplement the phase diagram of interacting hadronic matter with another method. As we know, the chiral transition line in quark models is usually derived with quark condensate $\phi$, which is directly related to the dynamical quark mass. The chiral crossover separation line can be derived with $\partial \phi/\partial \mu$ (or $\partial \phi/\partial T$) taking the extreme value for a given temperature~(chemical potential).
$\partial \phi/\partial \mu$ (or $\partial \phi/\partial T$) exhibits a jump for the first-order chiral transition.

\begin{figure}[htbp]
\begin{center}
\includegraphics[scale=0.35]{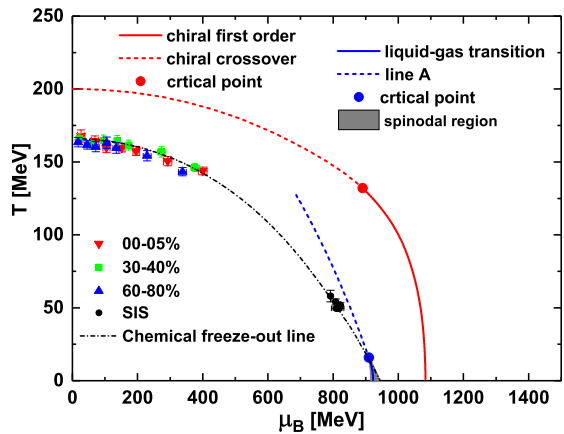}
\caption{\label{fig:2}Nuclear LG transition line and ``line A" ( where the dynamical nucleon mass varies most quickly), chiral transition line,  extracted chemical freeze-out temperature versus baryon chemical potential from the STAR experiments~\cite{Adamczyk17} and SIS data~\cite{Cleymans99,Becattini01,Averbeck03}, as well as chemical freeze-out curve fitted by Cleymans~{\it et~al.}\cite{Cleymans06}.   }
\end{center}
\end{figure}

Similarly,  the $\sigma$ field or the effective nucleon mass $m^*_N=m_N-g_\sigma \sigma$  in the interacting hadronic matter varies with the chemical potential and temperature.  Technically, we could perform a similar operation as in quark model.
 With the condition of $\partial \sigma/ \partial \mu_B$  (or $\partial m^*_N/ \partial \mu_B$)
 taking the extrema, we can also obtain a curve, as plotted in Fig.~\ref{fig:2} with the blue color. The blue solid line at low temperatures is just the nuclear LG transition line, but the blue dashed line, labeled line A, does not mean a phase transition. It only implies that the dynamical nucleon mass changes most quickly on the line, completely different from the chiral crossover transition in physics. 
 Whereas we will see that the blue line is closely related to the structure of density fluctuation induced by  hadronic interactions at low temperatures. For the convenience of  comparison with the quark chiral transition and the experimental data, we also plot in Fig.~\ref{fig:2} the chiral transition line derived in the Poyakov-Nambu--Jona-Lasinio model with $T_0=210\,$MeV~\cite{Shao2018} (The chiral transition line will move to higher temperatures and large chemical potentials if $T_0=270\,$MeV is taken) and the chemical freeze-out conditions derived from the STAR experiments \cite{Adamczyk17} and the SIS data~\cite{Cleymans99,Becattini01,Averbeck03}, as well as the fitted chemical freeze-out curve  in~\cite{Adamczyk17}.

\begin{figure}[htbp]
\begin{center}
\includegraphics[scale=0.35]{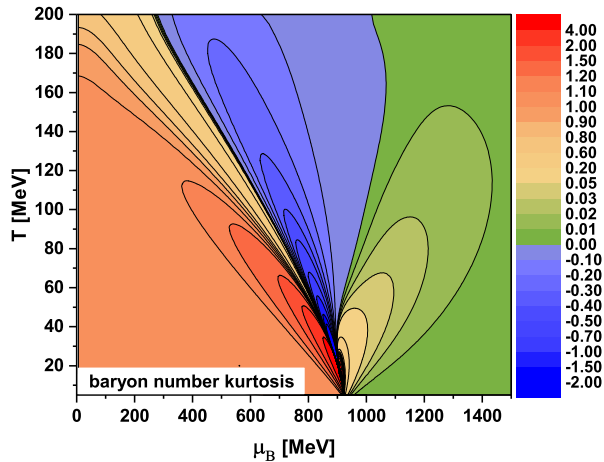}
\caption{\label{fig:3}Contour lines of net-baryon number kurtosis in the interacting hadronic matter. $\kappa \sigma^2$  is negative in the blue area and positive in the rest area. }
\end{center}
\end{figure}

\begin{figure}[htbp]
\begin{center}
\includegraphics[scale=0.34]{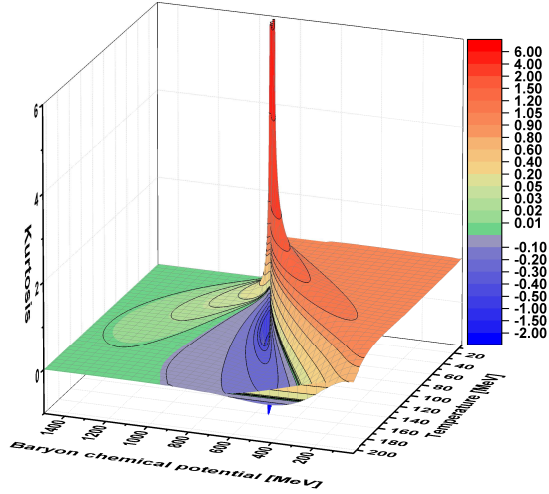}
\caption{\label{fig:4}Three-dimensional structure of  net-baryon number kurtosis in the interacting hadronic matter. $\kappa \sigma^2$ is negative~(positive) in the blue area~(in the rest region),  divergent at the critical end point of liquid-gas transition.  The blue dashed line~(line A) plotted in Fig.~\ref{fig:2}  lies in the  bottom of the valley in the blue area in the critical region.}
\end{center}
\end{figure}

In the following we demonstrate the density fluctuations induced by hadronic interactions. The  contour map  of net-baryon number kurtosis  $\kappa \sigma^2$ is plotted in Fig.~\ref{fig:3}. 
 The value of $\kappa \sigma^2$  is negative in the blue area and positive in the rest area. 
To visualize the fluctuation distribution, we also plot the three-dimensional landscape of $\kappa \sigma^2$ in Fig.~\ref{fig:4}. This figure clearly demonstrates the structure of net-baryon number kurtosis in the   $T-\mu_{B}$ phase diagram. It shows that   $\kappa \sigma^2$ diverges at the critical end point of the nuclear LG transition.  At the boundary of the first-order LG transition, $\kappa \sigma^2$ also takes relatively larger finite values. In particular,   in the critical region, the blue dashed line~(line A) given in Fig.~\ref{fig:2}    coincidentally lies in the  bottom of the valley in the blue area in Fig.~\ref{fig:3}  and Fig.~\ref{fig:4}.  All these indicate that the variation of  dynamical nucleon mass, i.e., the interaction between nucleons, plays a crucial role on the density fluctuations at low temperatures and large chemical potentials.  
It is easy to see that the topological structure of $\kappa \sigma^2$ induced by hadronic interactions at low temperature is  similar to the fluctuations induced by quark chiral transition~\cite{Shao2018}. 


To connect with heavy-ion collision experiments, we discuss possible signals in observations related to the chiral and nuclear liquid-gas transitions.
Firstly, the data based on the thermal statistical model~\cite{Adamczyk17,Cleymans06,Becattini2013}   
indicate that the chemical freeze-out temperatures with the collision energies  $\sqrt {s_{NN}}=7.7\sim27\,$GeV  are around $150\,$MeV. Therefore, the density fluctuation distributions at high temperature are mainly determined by quark chiral transition.
If the observed non-monotonic  behavior and large deviation from the baseline of  $\kappa \sigma^2$ with $\sqrt {s_{NN}}=7.7\sim27\,$GeV in BES~I can be confirmed in BES II, the existence of the chiral first-order transition with a CEP will be supported considerably. 

On the other hand, according to the thermal fit of SIS data with  $\sqrt {s_{NN}}=2.32\,$GeV,  $\mu_B\simeq820\,$MeV and $T\simeq50\,$MeV can be reached at chemical freeze-out~\cite{Cleymans99,Becattini01,Averbeck03}. The corresponding chemical freeze-out conditions are close to the critical region of nuclear LG phase transition, as shown in Figs.~\ref{fig:2}-\ref{fig:4},  and it happens that the fitted chemical freeze-out line~\cite{Cleymans06} enters into the spinodal region of nuclear liquid-gas transition. Therefore, with the decrease of collision energy,  the net-baryon number $\kappa \sigma^2$ possibly increases again in the critical region of nuclear LG transition.  As a matter of fact, as shown in Fig.~\ref{fig:4}, even not very close to the critical end point, the fluctuations will still be strengthened above the critical temperature.  It means that we can extract the information of nuclear LG transition even the phase transition is not triggered. 
The related signals will  be possibly observed in heavy-ion collision experiments with lower collision energies.  
Different from the previous methods, the measurement of the energy dependence of net-proton fluctuation distributions provides a new way to explore the nuclear LG phase transition. It is also helpful to ascertain the first-order chiral transition.

According to the critical behavior of  $\kappa \sigma^2$, the non-monotonic energy dependence of  $\kappa \sigma^2$ appears for each critical point. Therefore, if the chemical freeze-out line passes though the critical regions of the chiral and nuclear LG transitions, the measured net-proton number $\kappa \sigma^2$ as a function of collision energy
will has a double-peak structure. 
Although the realistic result are more complicated than the  ideal picture, it is anticipated that the essential feature can be retained.
In experiments, maybe  lower collision energies than those in the planed experiments are required to produce clear  signals at low temperatures.

\begin{figure}[htbp]
\begin{center}
\includegraphics[scale=0.35]{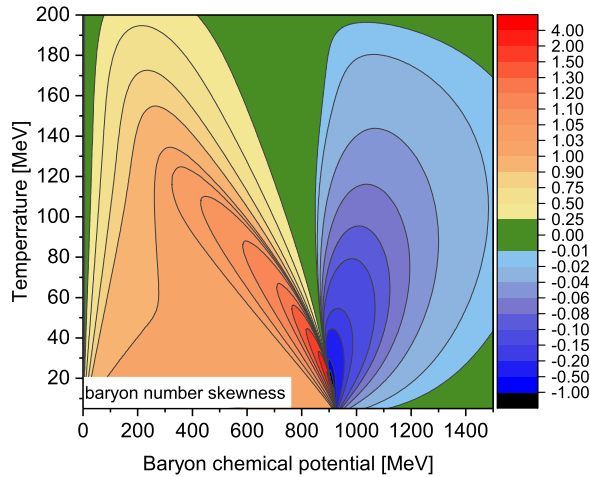}
\caption{\label{fig:5}Contour lines of  net-baryon number skenewss of the interacting hadronic matter. }
\end{center}
\end{figure}

\begin{figure}[htbp]
\begin{center}
\includegraphics[scale=0.35]{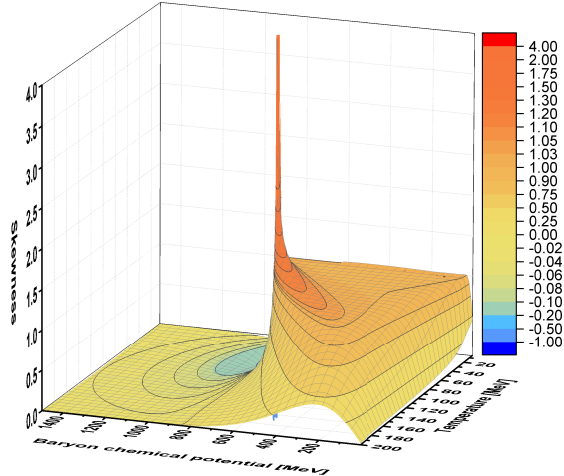}
\caption{\label{fig:6}Three-dimensional structure of  net-baryon number skewness of the interacting hadronic matter. $\kappa \sigma^2$ is negative~(positive) in the blue area~(in the rest region) and divergent at the critical end point.}
\end{center}
\end{figure}

Finally, we present the net-baryon number skewness $S\sigma$  in Fig.~\ref{fig:5} and Fig.~\ref{fig:6}.  Similar to the net-baryon number kurtosis, the two figures show that the skewness diverges at the critical end point. Below the critical temperature, skewness takes positive values on the gas phase of the nuclear LG transition, but on the liquid phase at high density it takes minus values. Above the critical temperature, the sign of skewness is separated by the blue dashed line~(line $A$) plotted in Fig.~\ref{fig:2} in the critical region. These features once again indicate that the interaction of hadronic matter plays a curial role on the density fluctuation distributions. As discussed above, the skewness will also probably demonstrate a double-peak structure as a function of collision energy.

To display the structure of  the fluctuation distributions induced by hadronic interactions, in Figs. \ref{fig:3}-\ref{fig:6},  the calculations from low temperatures to high temperatures up to 200 MeV are performed.  We should  note that only the values at low temperatures are reliable because the nonlinear Walecka model model with a few degrees of freedom of hadrons is only effective at  low temperatures. 
With the increase of temperature more hadronic degrees of freedom will appear, and then this model will gradually become invalid.  
Thus the calculation is only responsible for low-temperature results, corresponding to the collisions with relatively lower energies.
However, as the energy is lowered, the number of produced nucleons will decrease, disadvantageous to measure fluctuations or even no fluctuation can be generated at very low energies. Fortunately, for a first-order transition, the structure of fluctuation distributions exists in certain region. Even the nuclear LG transition is not directly triggered, it is still possible to extract the relevant signals according to the fluctuations above the critical temperature, which relaxes to a certain degree the requirement on collision energy.

\section{summary}
In summary,  we investigated the net-baryon number fluctuations induced by hadronic interaction after the hadronization of QGP at low temperatures. 
We found that hadronic interactions can also generate strong density fluctuations in the critical region of nuclear liquid-gas phase transition.

If the chemical freeze-out line passes by the critical regions of both the chiral and nuclear liquid-gas phase transitions, the measured  $ \kappa \sigma^2$ ($S\sigma$) of net-proton number as a function of  collision energy may present a double-peak structure. 
Even the nuclear liquid-gas phase transition is not triggered, the peak structure at low temperatures induced by  hadronic interactions may also be observed since the structure of fluctuation distributions exists in certain region. The strength of fluctuation signal depends on the distance to the critical end point. The available data suggest that lower collision energies are beneficial  to produce more clear signals at low temperatures.
If such a phenomenon can be observed in experiments, it will provide useful information to identify the phase structure of strongly interacting matter. In particular, the fluctuation distributions at low temperatures provide a new approach to investigate  hadronic interactions and nuclear liquid-gas phase transition.

\section*{Acknowledgements}
The authors would like to thank Prof. Yu-xin Liu and Wei-jie Fu for fruitful discussion.
This work is supported by the National Natural Science Foundation of China under
Grant No.~11875213 and the
Natural Science Basic Research Plan in Shaanxi Province of
China (Program No. 2019JM-050).

\end{document}